\documentclass[letterpaper]{article}
\usepackage[affil-it]{authblk}

\usepackage{amsmath}
\usepackage{amsthm}

\usepackage{caption}

\usepackage{algorithm}
\usepackage{algpseudocode}
\usepackage{float}
\usepackage{ifthen}
\usepackage{soul} 
\usepackage{tikz}
\usepackage[export]{adjustbox}
\usepackage{placeins}

\usepackage{fixltx2e}
\usepackage{upquote,textcomp}

\usepackage{palatino}
\usepackage{subcaption}
\usepackage{txfonts}
\usepackage{array}
\usepackage{calc}
\usepackage{pstricks}
\usepackage{graphicx}
\usepackage{epstopdf}
\usepackage{colortbl}
\usepackage{units}
\usepackage[makeroom]{cancel}

\usetikzlibrary{shapes, arrows}

\makeatletter

\newcommand\floatc@plainthin[2]{
\setbox\@tempboxa\hbox{{\@fs@cfont #1:} #2}
   \hbox to\hsize{\hfil\box\@tempboxa\hfil}\fi}
\newcommand\fs@plainthin{
  \def\@fs@cfont{\rmfamily}\let\@fs@capt\floatc@plainthin
  \def\@fs@pre{}
  \def\@fs@post{\vspace{-2.5em}}
  \def\@fs@mid{\vspace\abovecaptionskip\relax}
  \let\@fs@iftopcapt\iffalse}
\makeatother
\floatstyle{plainthin}
\newfloat{AlgFloat}{h}{lop}

\algnewcommand\algorithmicinput{\textbf{INPUT:}}
\algnewcommand\INPUT{\item[\algorithmicinput]}
\algnewcommand\algorithmicoutput{\textbf{OUTPUT:}}
\algnewcommand\OUTPUT{\item[\algorithmicoutput]}

\makeatletter
\@ifclassloaded{beamer}
{}
{

\newtheoremstyle{break}  
  {\topsep}   
  {\topsep}   
  {\itshape}  
  {-1em}      
  {\bfseries} 
  {}         
  {0pt}  
  {}          
\theoremstyle{break}

}
\makeatother

\usepackage[numbers,sort&compress]{natbib}
\graphicspath{{./epistasisSameGene/figures/}{./falconBioinfo/figures/}}

\usepackage{url}
\usepackage{fullpage}
\usepackage{setspace}

\makeatletter
\@addtoreset{algorithm}{section}
\makeatother

\title{Dynamic epistasis for different alleles of the same gene}

\date{\today}

\begin{document}

\newcounter{LinBrandonCoAuthor}
\author[1]{Lin Xu\footnote{contributed equally}%
\protect\setcounter{LinBrandonCoAuthor}{\value{footnote}}%
}
\newcommand\LinBrandonCoAuthorMark{\footnotemark[\value{LinBrandonCoAuthor}]}%

\author[2]{Brandon Barker\protect\LinBrandonCoAuthorMark}


\author[3,4]{Zhenglong Gu%
  \thanks{Electronic address: \texttt{zg27@cornell.edu}; Corresponding author}}

\affil[1]{Division of Hematology/Oncology, Department of Pediatrics,
    University of Texas Southwestern Medical Center, Dallas, TX, USA}
\affil[2]{Center for Advanced Computing,
    Cornell University, Ithaca, NY, USA.}
\affil[3]{Division of Nutritional Sciences, Cornell University,
  Ithaca, NY, USA.}
\affil[4]{Tri-Institutional Training Program in Computational
  Biology and Medicine, New York, NY, USA.}

\newboolean{thesisStyle}
\setboolean{thesisStyle}{true} 

\maketitle

                                       %
\newcommand{\captionpage}[2]{
\newpage
\vspace*{\fill}
\begin{center}
\captionof{#1}{#2}
\end{center}
\vspace{\fill}
\newpage
}

\tikzstyle{line} = [draw, very thick, color=black!50, -latex']
\tikzstyle{cloud} = [draw, ellipse,fill=red!20, node distance=2em]
\tikzstyle{decision} = [diamond, draw, fill=blue!20, align=center,
    text badly centered, node distance=6em, inner sep=0pt]
\tikzstyle{block} = [rectangle, draw, 
    text centered, align=center, node distance=4em]
\tikzstyle{iogram} = [trapezium, draw, fill=black!10, 
    trapezium left angle=70, trapezium right angle=-70,
    text centered, align=center, node distance=5em]
\tikzset{
  onslide/.code args={<#1>#2}{\only<#1>{\pgfkeysalso{#2}}}, 
}


\DeclareRobustCommand\suppOrApp{%
  \ifthenelse{\boolean{thesisStyle}}%
    {}%
    {Supplementary}%
}
\DeclareRobustCommand\Fig{%
  \ifthenelse{\boolean{thesisStyle}}%
    {Figure}%
    {Fig.}%
}

\DeclareRobustCommand\Figs{%
  \ifthenelse{\boolean{thesisStyle}}%
    {Figures}%
    {Figs.}%
}


\ifthenelse{\boolean{thesisStyle}}{%
  \floatstyle{plaintop}
  \restylefloat{table}
}
{}


\newcommand\D{\mathrm{d}}

\newcommand{\ANDw}{\land}
\newcommand{\ORw}{\lor}

\newcommand{\introSameGeneCredit}{%
\ifthenelse{\boolean{thesisStyle}}{%
  \footnote{This chapter is taken from material in \citealt{Shestov2013a}.
Brandon Barker is the primary author of all material found herein.}%
}%
{}%
}


\newcommand{\epiSameGeneCredit}{%
\ifthenelse{\boolean{thesisStyle}}{%
  \footnote{This chapter is published as \citet{Xu2012}.
Brandon Barker and Lin Xu contributed equally to this work.
It is additionally available in \citet[chapter 4]{Xu2012a}.}%
}%
{}%
}

\newcommand{\epiSameGeneAbstract}{%
Epistasis refers to the phenomenon in which phenotypic consequences
caused by mutation of one gene depend on one or more mutations at
another gene. Epistasis is critical for understanding many genetic and
evolutionary processes, including pathway organization, evolution of
sexual reproduction, mutational load, ploidy, genomic complexity,
speciation, and the origin of life. Nevertheless, current
understandings for the genome-wide distribution of epistasis are
mostly inferred from interactions among one mutant type per gene,
whereas how epistatic interaction partners change dynamically for
different mutant alleles of the same gene is largely unknown. Here we
address this issue by combining predictions from flux balance analysis
and data from a recently published high-throughput experiment. Our
results show that different alleles can epistatically interact with
very different gene sets. Furthermore, between two random mutant
alleles of the same gene, the chance for the allele with more severe
mutational consequence to develop a higher percentage of negative
epistasis than the other allele is 50-70\% in eukaryotic organisms,
but only 20-30\% in bacteria and archaea. We developed a population
genetics model that predicts that the observed distribution for the
sign of epistasis can speed up the process of purging deleterious
mutations in eukaryotic organisms. Our results indicate that epistasis
among genes can be dynamically rewired at the genome level, and call
on future efforts to revisit theories that can integrate epistatic
dynamics among genes in biological systems\epiSameGeneCredit.
}


\newcommand{\epistasisEnviroAbstract}{%
Epistasis describes the phenomenon that mutations at different loci do
not have independent effects with regard to certain
phenotypes. Understanding the global epistatic landscape is vital for
many genetic and evolutionary theories. Current knowledge for
epistatic dynamics under multiple conditions is limited by the
technological difficulties in experimentally screening epistatic
relations among genes. We explored this issue by applying flux balance
analysis to simulate epistatic landscapes under various
environmental perturbations. Specifically, we looked at gene-gene
epistatic interactions, where the mutations were assumed to occur in
different genes. We predicted that epistasis tends to become more
positive from glucose-abundant to nutrient-limiting conditions,
indicating that selection might be less effective in removing
deleterious mutations in the latter. We also observed a stable core of
epistatic interactions in all tested conditions, as well as many
epistatic interactions unique to each condition. Interestingly, genes
in the stable epistatic interaction network are directly linked to
most other genes whereas genes with condition-specific epistasis form
a scale-free network. Furthermore, genes with stable epistasis tend to
have similar evolutionary rates, whereas this co-evolving relationship
does not hold for genes with condition-specific epistasis. Our
findings provide a novel genome-wide picture about epistatic dynamics
under environmental perturbations.
}

\newcommand{\epistasisEnviroAuthorSummary}{%
Epistasis, often referred to as genetic interactions, occur when
mutational effects of genes depend on each other. Aside from often
times complicating the way in which the phenotype of an organism
relates to its genotype, epistatic interactions (or epistases) are
essential to several important theories in biology, especially in
evolution. Due to the difficulty in experimentally assessing epistasis
across an entire genome, we employed mathematical modeling of the
metabolic network of baker's yeast to comprehensively simulate genetic
interactions for virtually all known metabolic genes in the
organism. We performed comprehensive simulations in 17 different
environments, which differ by their nutrients. We characterized a
trend that occurs in genetic interactions when yeast is transferred
from a glucose-abundant environment to other environments. We also
found that both the set of genetic interactions present in all
conditions and the set of interactions present in a single environment
are fairly large sets with highly different connectivity. Furthermore,
the set present in all conditions tends to consist of gene pairs with
similar evolutionary rates.
}


\newcommand{\falconAbstractMotivation}{%
A major theme in constraint-based modeling is unifying 
experimental data, such as biochemical information about the reactions
that can occur in a system or the composition and localization of enzyme
complexes, with high-throughput data including expression data,
metabolomics, or DNA sequencing. The desired result is to increase
 predictive capability and improve our understanding of metabolism.
 The approach typically employed when only gene (or protein) intensities
are available is the creation of tissue-specific models, which reduces
the available reactions in an organism model, and does not provide an
objective function for the estimation of fluxes.
}

\newcommand{\falconAbstractResults}{%
We develop a method, flux assignment with LAD (least absolute
deviation) convex objectives and normalization (FALCON),
 that employs metabolic network reconstructions along with expression
data to estimate fluxes. In order to use such a method, accurate
measures of enzyme complex abundance are needed, so we first
present an algorithm that addresses quantification of complex
abundance. Our extensions to prior techniques include the
capability to work with large models and significantly improved
run-time performance even for smaller models, an improved analysis of
enzyme complex formation, the ability to handle large enzyme
complex rules that may incorporate multiple isoforms, and either
maintained or significantly improved correlation with experimentally
measured fluxes.
}

\newcommand{\falconAbstractAvail}{%
FALCON has been implemented in MATLAB and ATS, and can be downloaded
from: \url{https://github.com/bbarker/FALCON}. ATS is not required to
compile the software, as intermediate C source code is available. 
FALCON requires use of the COBRA Toolbox, also implemented in MATLAB.
}


\newcommand{\epiBeneMutAbstract}{%
Existing literature has only dealt with the simulation of strictly
deleterious mutants rather than beneficial mutants in the
constraint-based modeling literature, which is chiefly due to existing
studies optimizing the fitness function, which leaves no room for
improvement. In this study, we develop a constraint-based approach
that can simulate beneficial, neutral, and deleterious mutations.  We
show that this simulation technique can be useful for understanding
adaptive trajectories, and we develop a software library for the
analysis of evolutionary paths.  Our mechanistic model can reproduce
the distribution of
epistases between beneficial mutations that was observed in a data set
and a population genetic model fit to the same data set, showing that
our model behaves appropriately in this conext and may be a useful
tool for further evolutionary analyses. Finally, in experimental data
sets and in our simulations, slightly beneficial mutations are much
more likely to have positive (synergistic) epistasis with other
beneficial mutations, making their likelihood of becoming fixed higher
than would be expected without considering epistatic effects.
}

\captionsetup{labelfont=bf}




  %
                                       %

\begin{abstract}
\epiSameGeneAbstract
\end{abstract}

\def\suppOrApp{}

\section{Introduction}

Epistasis between two deleterious mutations is positive when a double
mutant causes a weaker mutational defect than predicted from
individual deleterious mutations, and is negative when the double
mutant causes a larger defect \citep{Phillips2008, Boone2007}. In a
population with sexual
reproduction, positive epistasis alleviates the total harm when
multiple deleterious mutations combine together and thus reduces the
effectiveness of natural selection in removing these deleterious
mutations, whereas negative epistasis can lower average mutational
load by efficiently purging deleterious mutants \citep{Kimura1966}. As
a consequence,
selective elimination of deleterious mutations would be especially
effective if negative epistasis is prevalent. It is important to
understand the distribution of epistasis among mutations, which plays
a central role in genetics and theoretical descriptions for many
evolutionary processes \citep{Phillips2008, Boone2007}.

Tremendous efforts have been put into genome-wide measurements for the
sign and magnitude of epistasis among different genes in various
species \citep{Costanzo2010, Tong2004, Pan2004, Pan2006,
Measday2002, Measday2005, Sopko2006, Collins2007,
Kornmann2009, Fiedler2009, Bonhoeffer2004, Roguev2008}.  A series of
high-throughput experimental platforms have been developed, such as
synthetic genetic array (SGA; \citealt{Costanzo2010, Tong2004}),
diploid-based synthetic lethality analysis with microarrays
\citep{Pan2004, Pan2006}, synthetic dosage-suppression and lethality
screen \citep{Measday2002, Measday2005, Sopko2006}, and
epistatic miniarray profiles \citep{Collins2007, Kornmann2009,
Fiedler2009}. The epistatic relations in these
experiments were mostly measured based on one mutant type (deletion
mutant) per gene. Few studies constructed multiple mutant alleles for
single genes to examine the dynamics of epistatic relations among
genes under different genetic perturbations. As a consequence, the
global landscape of epistasis for different alleles of the same gene
remains largely uninvestigated.

We address this issue by exploring epistatic differences among alleles
in the same gene for a large part of the genome by combining
experimental data with mathematical modeling using flux balance
analysis (FBA). FBA involves
the optimization of cellular objective functions and allows prediction
of in silico flux values and/or growth \citep{Mo2009, Becker2007,
Smallbone2009a}. FBA has been used to investigate the fitness
consequence of single-deletion mutants \citep{Papp2004, Ibarra2002}
and epistatic relations between metabolic reactions, genes, and
functional modules \citep{Harrison2007, Deutscher2006, Segre2005,
He2010}. The FBA predictions show good agreement with genome-wide
experimental studies \citep{Edwards2001, Segre2002_sb2013, Shlomi2005,
AbuOun2009, Durot2009, Feist2008, Fong2005a, Fong2005}.  One essential
advantage
of FBA modeling is that it can simulate epistasis between genes based
on different genetic mutants. Using this platform, together with data
from a recently published experiment \citep{Costanzo2010}, we were able to show that
epistasis can be rewired among genes, and that the sign of epistasis
can change dramatically at the global scale, depending on the mutant
alleles involved in the processes. Our study provides a genome-wide
picture on the dynamic epistatic landscape of various mutant alleles
for the same gene.

\section{Results}

\subsection{Epistatic Relations Between Genes Are Largely
Allele-Specific.}

\begin{figure}[h!]
\centering
\includegraphics[height=0.85\textheight]{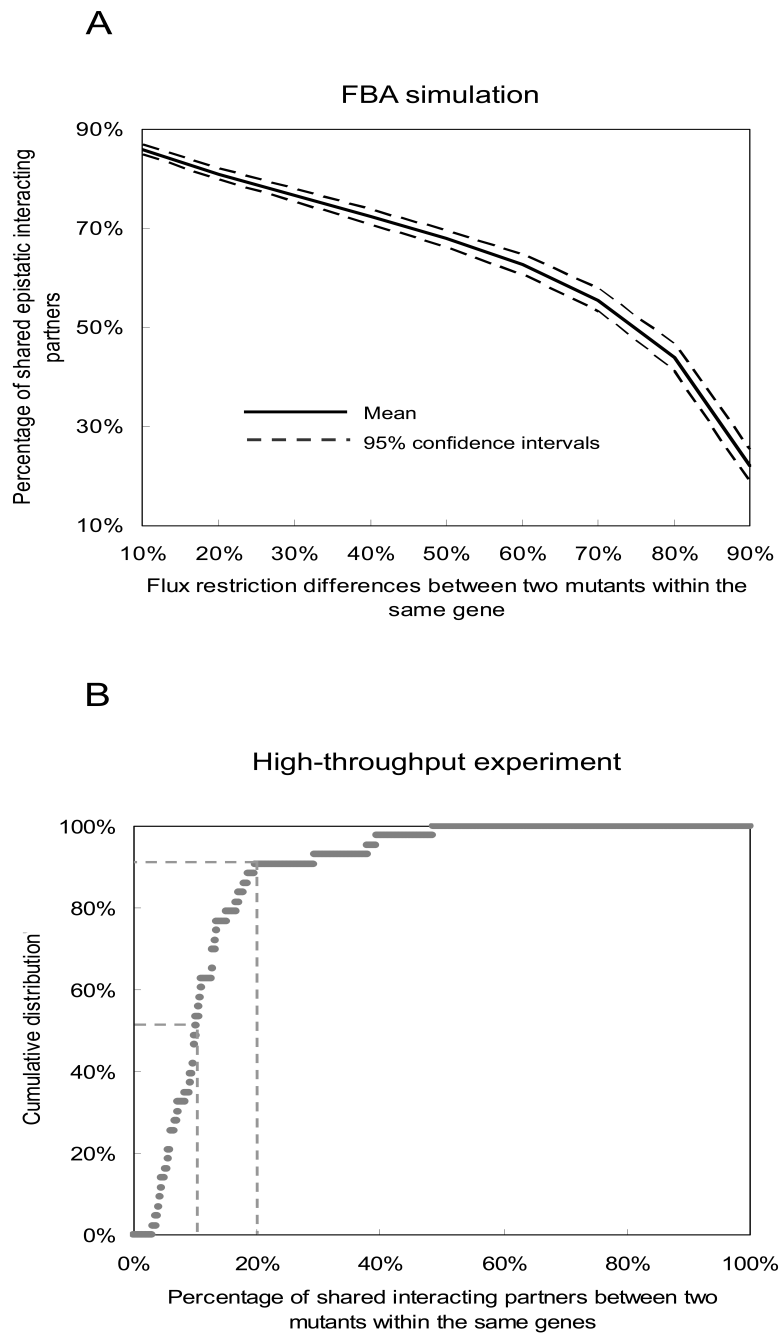}
\caption{Epistatic relations between genes are allele specific. (A)
FBA simulation results for the distribution of the percentage of
shared epistatic interaction partners between two mutant alleles
within the same gene. Solid and broken lines represent mean and 95\%
confidence intervals, respectively. (B) The cumulative distribution
for the percentage of shared epistatic interaction partners between
two mutant alleles within the same gene based on real experimental
data. Two broken lines represent 10\% and 20\% of shared epistatic
profiling, respectively.}
\label{fig:alleleSpecific}
\end{figure}

\begin{figure}[!htb]
\centering
  \includegraphics[width=\textwidth]{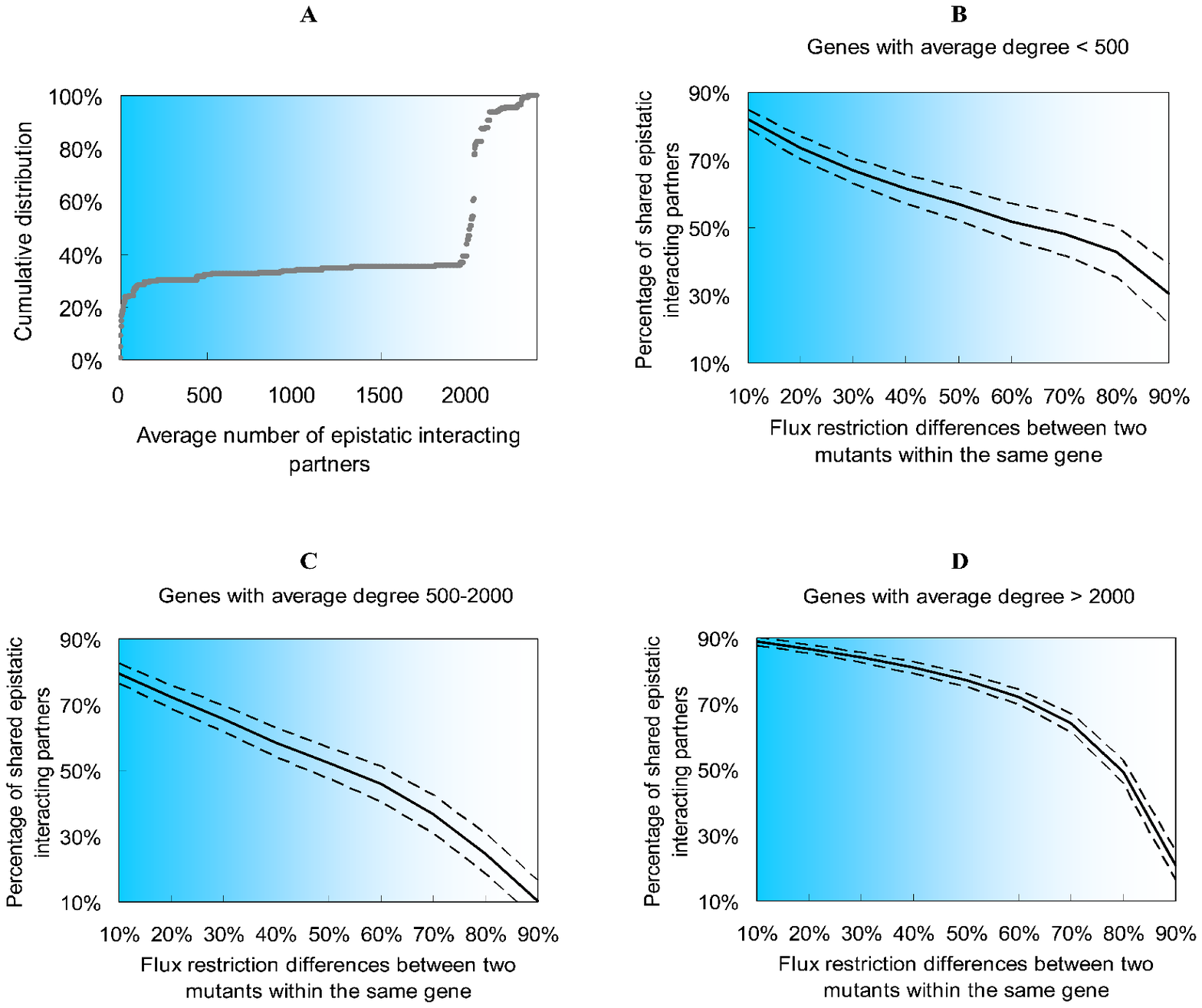}
\caption{The conclusion in Fig. 1 is not dependent on the average
number of epistatic interaction partners per gene. (A) The
distribution of average number of epistatic interaction partners per
gene. For each gene with epistasis, its average number of epistatic
interaction partners was calculated among all mutant alleles of this
gene. (B-D) A similar conclusion to that of Fig. 1 can be obtained
when we only use genes with fewer than 500 (B), 500-2,000 (C), and
more than 2,000 (D) average epistatic interaction partners. The same
methods in Fig. 1 were used here to generate B-D.}
\label{fig:sgeS1}
\end{figure}

\begin{figure}[!htb]
\centering
  \includegraphics[width=\textwidth]{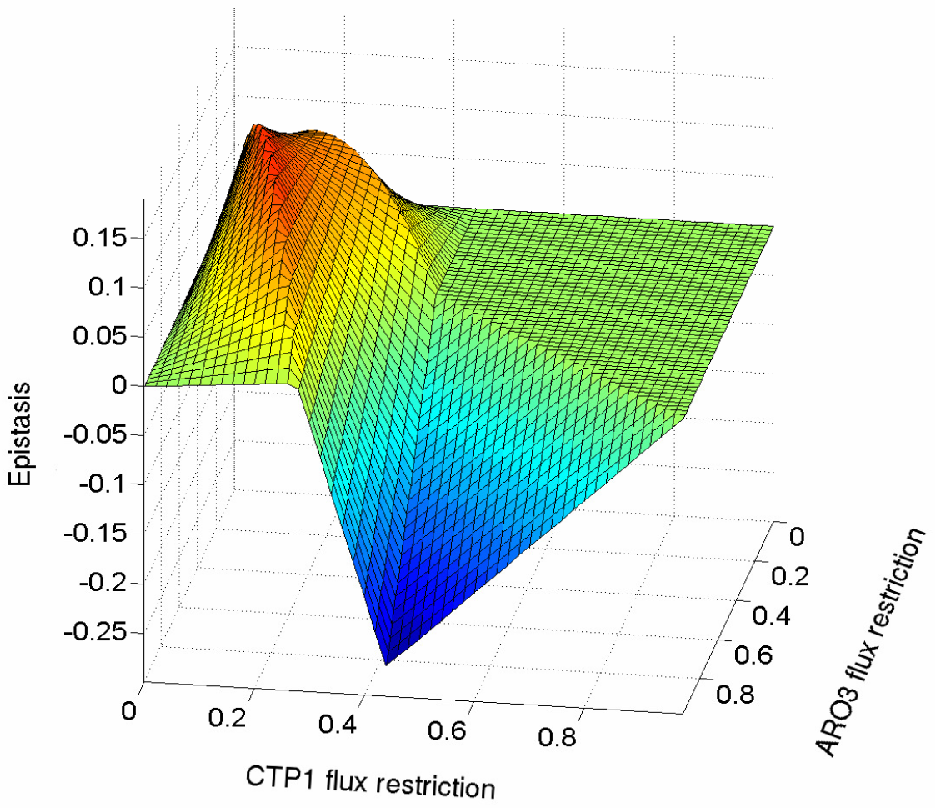}
\caption{A complex epistatic landscape exhibits a transition from
large positive to large negative epistasis values, along with a region
of zero epistasis. Epistasis is viewed as a function of the CTP1 and
ARO3 genes’ flux restriction. The color corresponds to the z-axis
(epistasis), with red being more positive, green being near zero, and
blue being more negative.}
\label{fig:sgeS2}
\end{figure}

\begin{figure}[!htb]
\centering
  \includegraphics[width=\textwidth]{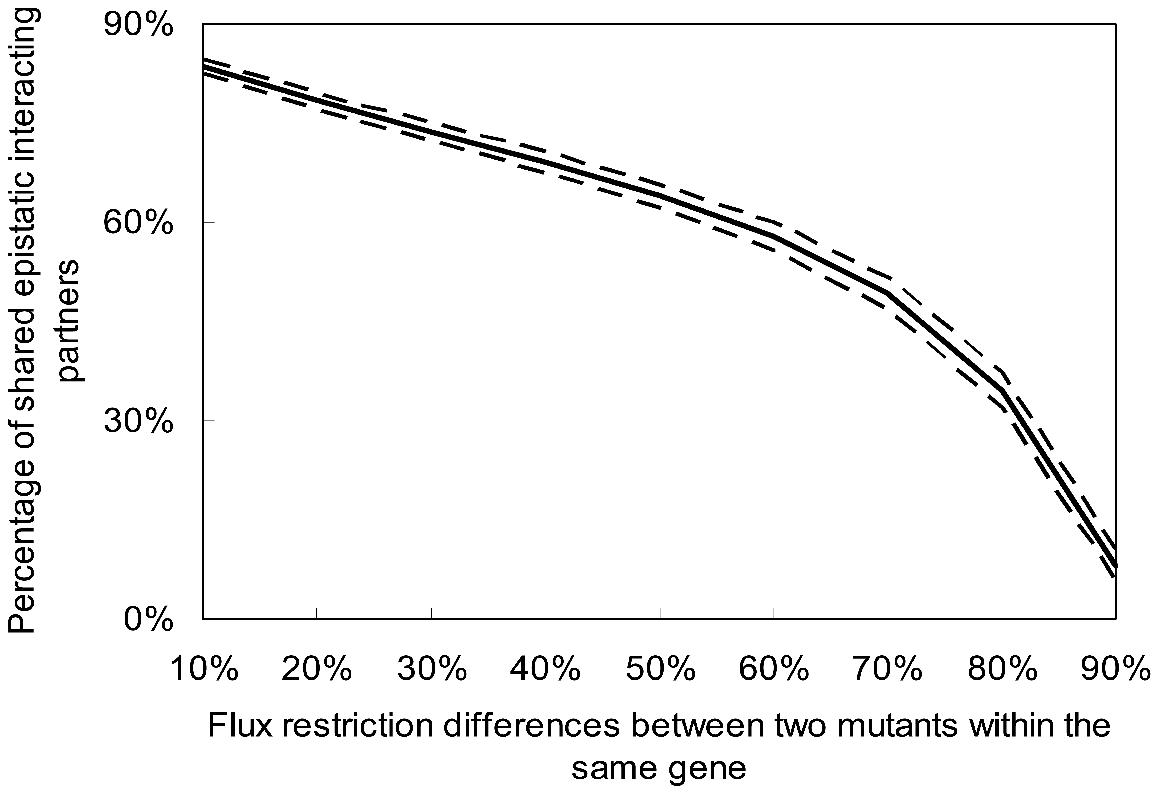}
\caption{Percentage of shared epistatic interacting partners based on
flux differences between two mutant alleles of the same gene. The
analysis procedure is the same as Fig. 1A, but instead of using the
S. cerevisiae model, here we repeated the analysis using the E. coli
model (38).}
\label{fig:sgeS3}
\end{figure}

We first used the yeast \textit{Saccharomyces cerevisiae} metabolic
reconstruction iMM904 \citep{Mo2009} to examine the distribution of epistasis
under various genetic mutant alleles. The reconstruction is a
genome-scale metabolic model, having 904 metabolic genes associated
with 1,412 reactions. For each gene, we simulated genetic
perturbations that retain the corresponding flux from 90\% to 0\% in
decrements of 10\% of its WT (optimal) flux. As a result, 10 different
single mutants per nonessential gene and nine different single mutants
per essential gene (the 0\% flux mutants in these genes represent
lethal deletion for which epistasis cannot be calculated) were
simulated. We computed the fitness of the single mutants and double
mutants with any possible pairwise allele combination of different
genes. These data were used to infer the epistatic relationships among
genes. In total, over 40 million simulations were conducted.  To
investigate the dynamics of epistasis among genes, we calculated the
percentage of shared epistatic interaction partners between any two
mutants within the same gene. Two mutant alleles are defined to share
an epistatic interaction partner (a mutant from another gene) if they
both epistatically interact with this mutant and the signs of
epistasis are the same. The percentage of shared epistatic interaction
partners between two mutants is calculated as the number of their
shared epistatic interaction partners divided by the sum of their
total epistatic interaction partners. As shown in
\Fig~\ref{fig:alleleSpecific}A, our results
indicate that the percentage of shared epistatic interaction partners
between two mutants of the same gene decreases as the flux difference
between them increases. Two mutants of the same genes could have as
low as only about 20\% overlap between their epistatic interaction
partners, indicating that the epistatic profile of a gene is largely
dependent on the mutant types used. Our results also show that the
average number of epistatic interaction partners per gene do not
affect this conclusion (\Fig~\ref{fig:sgeS1}). Interestingly, there
are cases where the sign of epistasis between two genes can even
change under varying mutant types (an example is in
\Fig~\ref{fig:sgeS2}, and all pairs with reversed sign of epistasis
are listed in Dataset S1). However, such events are rare
(about 1.2\% of all gene pairs that show epistatic
interactions). Furthermore, we repeated the above FBA analysis for
another species, \textit{Escherichia coli}, and the results confirmed
the above trend (\Fig~\ref{fig:sgeS3}).



In a recently released high-throughput experiment that measured
genome-wide epistatic relations among genes in \textit{S. cerevisiae} \citep{Costanzo2010},
there were 43 mutant pairs having two different mutant alleles of the
same gene (Dataset S2), each of which were experimentally crossed with
3,885 array gene deletion mutants to explore their epistatic relations
in the genome. In total, over 200,000 double mutants were
experimentally constructed. This dataset provides the most
comprehensive experimental source for investigating the epistatic
landscape of different mutant alleles in the same
gene. \Fig~\ref{fig:alleleSpecific}B shows
the empirical cumulative distribution for the percentage of shared
interaction partners between mutant pairs within the same gene. Our
results indicate that more than 50\% of mutant pairs within the same
gene have less than 10\% overlap of their epistatic interaction
partners, and about 90\% mutant pairs have less than 20\% overlap
(\Fig~\ref{fig:alleleSpecific}B). As shown in Dataset S2, the
functions of genes used in the
experiments are very diverse, and not restricted to metabolic
functions as genes in the FBA model. Nevertheless, the result from
experimental studies confirms our FBA modeling prediction that
different mutant alleles of the same gene can have very distinct
epistatic interaction partners in the genome. In addition, the
conclusions are robust under various epistasis thresholds
(\Fig~\ref{fig:sgeS4}).

\subsection{Sign of Epistasis for Individual Genes Depends on Mutation
Severity.}

The relative prevalence of positive vs. negative epistasis is of
tremendous importance for understanding many evolutionary processes
\citep{Phillips2008, Boone2007, Kimura1966}. 
In the following we addressed this issue for different alleles
of the same gene. Based on the above high-throughput experimental
dataset, we calculated the percentage of negative epistasis for each
mutant, defined as the number of negative epistatic partners for this
mutant divided by the overall number of its epistatic partners. We
then compared the percentage of negative epistasis between different
mutant alleles of the same gene in the experiment. Among 43 mutant
pairs in the study, 35 mutant pairs have significantly different
fitnesses between two mutants of the same gene. As shown in
\Fig~\ref{fig:expVerifyYeast1}A left, 21 mutant pairs (60\%) show that
alleles with more severe
defects have a higher chance than the other allele in the same gene to
develop negative epistasis in the genome.

\captionpage{figure}{%
Mutant alleles in the same gene with more severe defects tend
to have a higher percentage of negative epistasis in yeast. (A) The
two matrices represent all mutant pairs identified in real
experimental data (left) and FBA simulation (right) (fitness
difference $\left|\Delta f\right| \geq 0.01$; epistasis threshold
$\left|\epsilon\right| \geq 0.01$). Each cell
represents one mutant pair within the same gene. The color bar to the
right represents the normalized percentage of negative epistasis for
the mutant allele with more severe defects (percentage of negative
epistasis for the mutant allele with more severe defects divided by the sum of
percentage of negative epistasis for two mutant alleles). Red and
yellow colors represent that mutant allele with more severe defects in
the same gene has higher and lower percentage of negative epistasis
than the other allele, respectively. (B) Distribution for the number
of mutant pairs among randomly selected 35 pairs where mutants with
more severe defects have higher percentage of negative epistasis. The
arrow represents the observed number for the mutant allele pairs
within the same genes. (C) The percentage of mutant pairs in which the
mutant allele with more severe defects in the same gene has a higher
percentage of negative epistasis under various fitness difference and
epistasis thresholds during FBA simulations.}
\label{fig:expVerifyYeast1}

\begin{figure}[!htb]
\centering
\includegraphics[height=0.95\textheight]{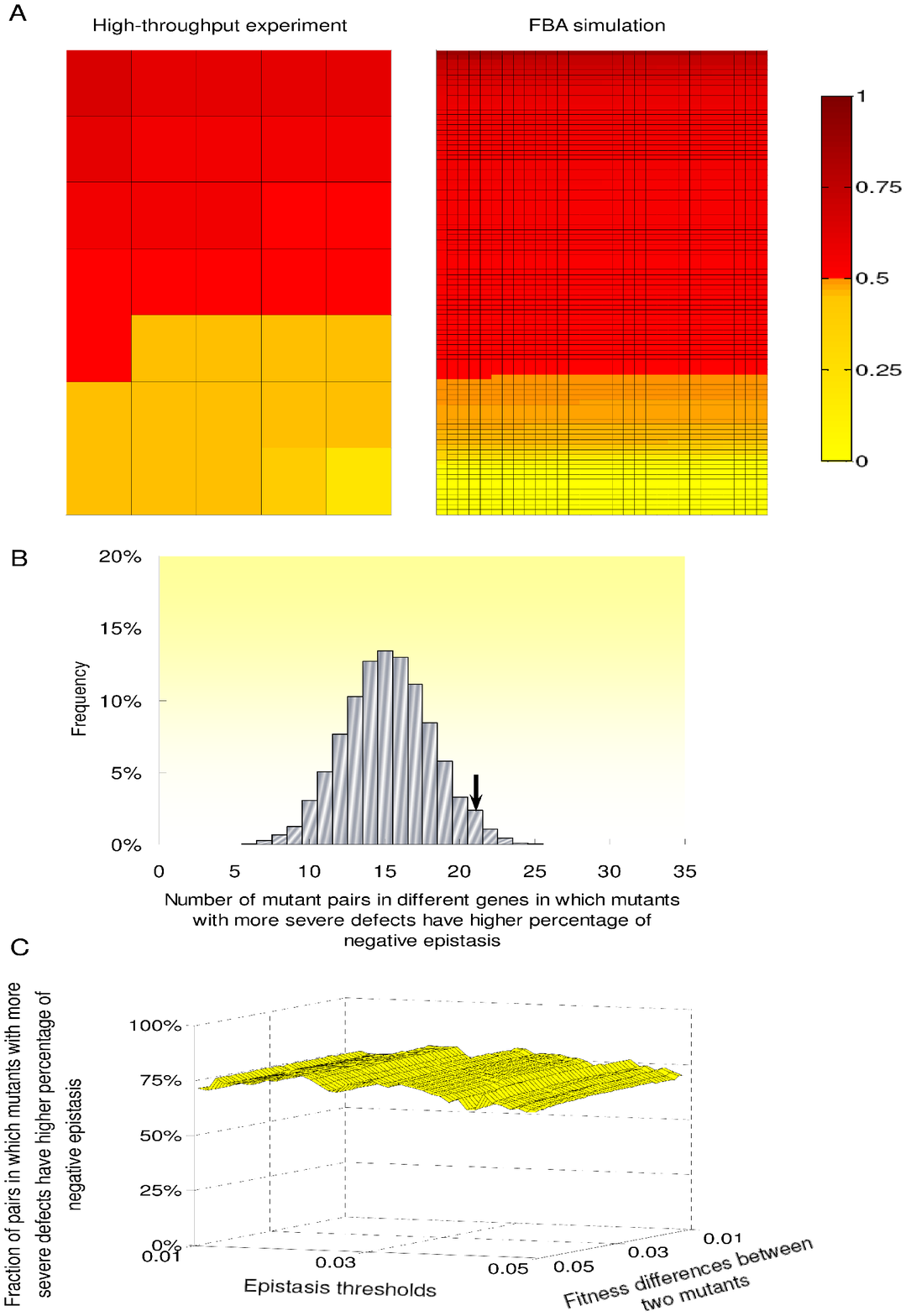}
\end{figure}

To see if this result could be caused by a systematic trend in the
high-throughput experiments, we randomly selected 35 pairs of mutants
from distinct genes that have the same fitness level for
single-deletion mutant and fitness difference between two mutants as
the above 35 pairs of mutants within the same genes, and compared
their relative prevalence of negative epistasis. The permutation was
repeated 100,000 times, and the result is depicted in
\Fig~\ref{fig:expVerifyYeast1}B. Among all repeats of randomly
selected 35 mutant pairs, only a small
percentage (4.1\%) have 21 or more mutant pairs where the mutant with
more severe defects has a higher chance than the other mutant to
develop negative epistasis in the genome, indicating that our
observation for different mutant alleles of the same gene is not
likely caused by the overall pattern in the high-throughput
experiments.

Using results from the above FBA simulation, we also confirmed the
same pattern that between mutant alleles of the same gene, the mutant
allele with more severe defect is more likely than the other allele to
develop negative epistasis in the genome
(\Fig~\ref{fig:expVerifyYeast1}A, right). Indeed, an even higher
percentage of mutant allele pairs in the FBA simulation
(about 70\%) than in real experiments (60\%) support this
conclusion. To avoid possible bias from the definition of epistasis
and fitness differences between mutant alleles in the FBA simulation,
we repeated the calculations based on multiple criteria and our
conclusion remains the same (\Fig~\ref{fig:expVerifyYeast1}C).

\begin{figure}
\centering
\includegraphics[width=\textwidth]{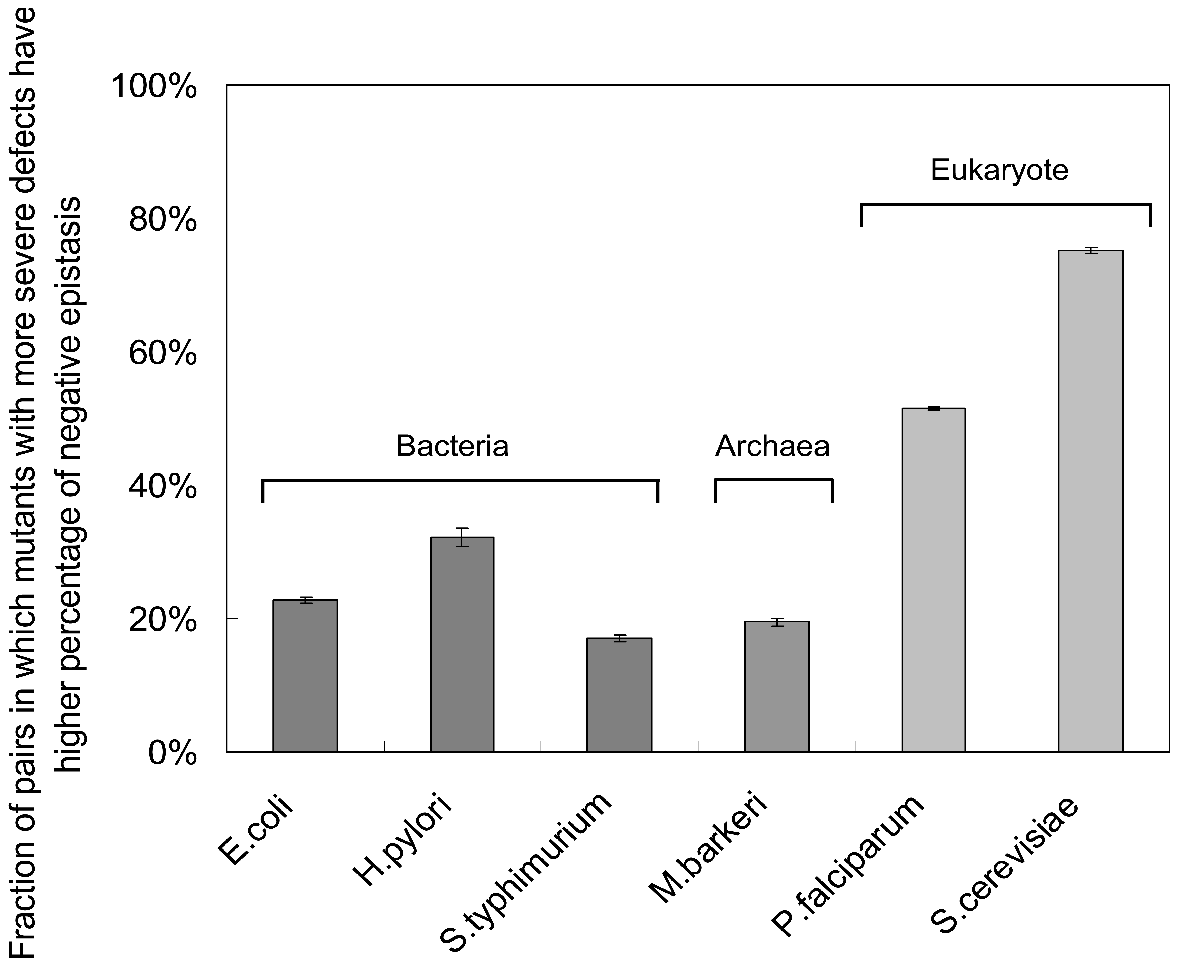}
\caption{Mutant alleles with more severe defects tend to have a higher
percentage of negative epistasis in eukaryotes than bacteria and
archaea. The y axis shows the percentage of mutant pairs in which
mutant alleles with more severe defects in the same gene have a higher
percentage of negative epistasis than the other allele. FBA
simulations were conducted for three bacterial species
(\textit{E. coli}, \textit{S. typhimurium}, and \textit{H. pylori}),
one archaea species (\textit{M. barkeri}), and two single-cell
eukaryote species (\textit{P. falciparum} and
\textit{S. cerevisiae}). The mean and SEs were based on results from
40 epistasis threshold values ranging from 0.01 to 0.05.}
\label{fig:speciesPurging}
\end{figure}

Our observation is surprising given that previous results based on
virus models or gene network simulations proposed a totally opposite
pattern at the genome level, i.e., mutations with larger mutational
defects are more likely to develop positive epistasis \citep{Burch2004, 
You2002, Sanjuan2006a, Azevedo2006, Lohaus2010}. We
further used the FBA simulations to explore the dynamics of epistasis
for various mutant alleles of the same gene in different
species. High-quality genome-wide metabolic networks in three bacteria
(\textit{Escherichia coli} \citep{Feist2007}, \textit{Salmonella typhimurium} \citep{Thiele2011},
and \textit{Helicobacter pylori} \citep{Thiele2005}), one archaea
(\textit{Methanosarcina barkeri} \citep{Feist2006}), and another single-cell
eukaryote (\textit{Plasmodium falciparum} \citep{Plata2010}) were used in our
simulation. As shown in \Fig~\ref{fig:speciesPurging}, when two mutant alleles of the same
gene are compared, in 22\%, 17\%, 32\%, and 19\% of cases for
\textit{E. coli}, \textit{S. typhimurium}, \textit{H. pylori}, and
\textit{M. barkeri}, respectively, mutant
alleles with more severe defects display higher percentages of
negative epistasis than the other allele, indicating that more
deleterious mutant alleles in the same gene indeed tend to develop
positive epistasis in these species. However, these numbers are
significantly smaller than that of yeast and another eukaryotic
organism, \textit{P. falciparum} (52\%). The conclusion is robust
under various epistasis thresholds (\Fig~\ref{fig:sgeS5}).

\subsection{Self-Purging Mechanism for Deleterious Mutations at the
Population Level}

Our above results indicate that between two random mutant alleles of
the same gene, the chance for the allele with more severe mutational
consequence to develop a higher percentage of negative epistasis than
the other allele is 50-70\% in eukaryotic organisms, but only 20-30\% in
bacteria and archaea. In other words, mutant alleles with more severe
defects in the same gene might have a higher chance to develop
negative epistasis in eukaryotic organisms than in bacteria and
archaea. We constructed a simple population genetic model as in
\Fig~\ref{fig:popgenPurging}A to address the evolutionary significance of this
observation. The genetic system has two genes: a query gene A, which
contains three different alleles (A\textsuperscript{S}: mutants with
severe defects; A\textsuperscript{D}: mutants with weak defects;
A\textsuperscript{WT}: WT), and a gene X, which has two different
alleles (mutant, X\textsuperscript{M}, and WT,
X\textsuperscript{WT}). We simulated the ratio of allele frequency
between the severe and the weak mutant alleles in
gene A under different probabilities of having negative epistasis
between these two alleles and the mutant allele in the gene X.

\captionpage{figure}{%
Increased efficiency of purging deleterious mutations in
eukaryotic organisms. (A) The population genetics model for allele
frequency changes from generation to generation. In the figure, $\rho$
and $\omega$ represent allele frequency and fitness, respectively. A
and X are genes with different alleles, and $\epsilon$ is the epistasis term
between mutant types of different genes. (B) The ratio of the severe
to the weak alleles of the A gene in the 50th, 100th, 150th, 200th,
250th, and 300th generations. Colors represent the ratio as indicated
at the bottom. The diagonal line in each panel represents the
situation where the severe and the weak mutant alleles have the same
probability of having negative epistasis in the genome. It is
noteworthy to point out that in each panel the ratio of the severe to
the weak alleles decreases, indicating increased efficiency of purging
the severe mutant allele, from the upper right (region I, the weak
mutant has more negative epistasis) to the bottom left (region II, the
severe mutant has more negative epistasis) part of the panel. The
arrows A and B are discussed in the text.}
\label{fig:popgenPurging}

\begin{figure}[!htb]
\centering
\includegraphics[width=0.95\textwidth]{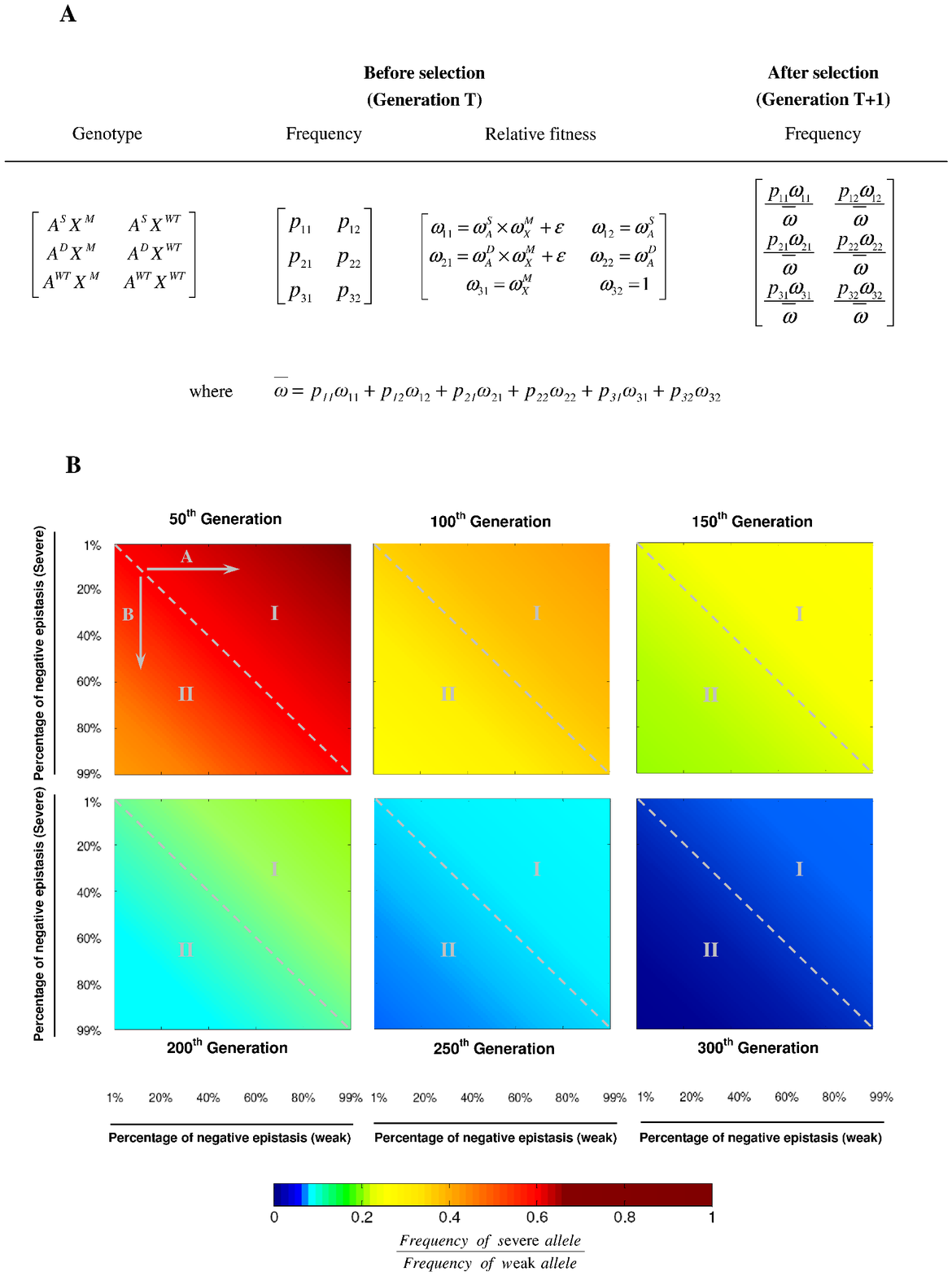}
\end{figure}

Our results in \Fig~\ref{fig:popgenPurging}B depict the simulation
results. The six panels in the figure represent the ratio of
A\textsuperscript{S} to A\textsuperscript{D} alleles in the 50th,
100th, 150th, 200th, 250th, and 300th generations, respectively. Our
simulations indicate that if the percentage of negative epistasis for
the severe mutant is kept as a constant, as the percentage of negative
epistasis for the weak mutation increases (as shown by the arrow A),
the ratio of the severe to the weak allele frequency would
increase. However, this ratio would decrease, indicating a faster
removal of the severe mutants from the population, in another
direction (as shown by the arrow B), i.e., the percentage of negative
epistasis for the weak mutant is kept as a constant, but the
percentage of negative epistasis for the severe mutant
increases. Therefore, the distribution for the sign of epistasis among
different alleles of the same gene observed in this study might
represent an efficient way for eukaryotic organisms to purge
deleterious mutations from populations.

\section{Discussion}
Our study represents a genome-wide theoretical survey for the dynamics
of global epistatic effects under various mutant alleles of the same
gene. We show that the epistatic profiling of a gene at the genome
level is largely dependent on mutant types involved. Our results
indicate that previous conclusions inferring epistatic relations among
genes based on only one mutant type per gene can be greatly improved
by using multiple mutant alleles. More importantly, our study shows
that mutant alleles with severe defects have a higher chance to
develop negative epistasis in eukaryotic organisms than in bacteria
and archaea. It has been speculated that eukaryotic organisms might
have more negative epistasis due to their increased complexity over
prokaryotic organisms \citep{Sanjuan2006, Sanjuan2008}. Even if this hypothesis is true,
however, our results for different mutant alleles of the same gene
cannot be directly inferred from this complexity argument.

Even though the mechanism underlying our observation remains to be
determined, we argue that such distributions for negative epistasis
among different alleles of the same genes have significant
evolutionary consequences, as shown in our population genetics
simulations (\Fig~\ref{fig:popgenPurging}). The origin and maintenance of sexual
reproduction remains one of the central issues in evolutionary
biology. Population genetics models have been proposed to explore the
impact of epistasis on the maintenance of sexual reproduction
\citep{Kondrashov1988, Otto2007, DeVisser2007, Kouyos2007}. The
mutational deterministic hypothesis posits that sex
enhances the ability of natural selection to purge deleterious
mutations by bringing them together into single genome through
recombination \citep{Kondrashov1988}. This explanation requires the prevalence of
negative epistasis at the genome level. Here we found that the
mutations with larger deleterious defects within the same gene have a
higher chance to develop negative epistasis in eukaryotic organisms
than bacteria and archaea. The model we proposed in
\Fig~\ref{fig:popgenPurging}, which is
based on the population genetics theory from Kondrashov \citep{Kondrashov1988},
indicates that such distribution of negative epistasis among different
alleles of the same gene in eukaryotic organisms might lead to more
efficient purging of deleterious mutations from populations, thus
providing a previously unappreciated evolutionary advantage for sexual
reproduction. We emphasize that these findings do not necessarily
provide sufficient evidence to explain the cause for the emergence of
sexual reproduction during evolution.

Although we found several unique characteristics regarding the global
epistatic landscape of different mutant alleles in the same gene,
three caveats need to be addressed. First, the FBA modeling used in
this study, which has been successfully applied to various research
problems \citep{Papp2004, Ibarra2002, Harrison2007, Deutscher2006,
Segre2005, He2010}, includes only metabolic genes in the
simulation. However, results from our analysis on the experimentally
defined epistatic relations among roughly 0.2 million double mutants
comprising about 4,000 \textit{S. cerevisiae} genes, which nearly
represent all functional categories in the budding yeast, confirmed
our major FBA modeling predictions.

Second, even though FBA is one of the most comprehensive computational
tools for simulating epistatic interactions among genes, there are
still many aspects that can be improved to aid in capturing the full
set of empirical genetic interactions \citep{Covert2001}. For example, rules for
transcriptional regulation and physical interactions can be integrated
into the current FBA framework to improve its accuracy \citep{Szappanos2011}. In
addition, mapping between individual alleles and metabolic flux
reduction is a complex process and difficult to measure experimentally
\citep{Banta2007}. It is noteworthy that in our simulations we have uniformly
evaluated fitness consequence based on the percentage of WT flux
attainable in a specific background. Depending on the regulation
dynamics of individual genes, such uniform sampling may be unlikely to
correspond to random sampling of mutant alleles. For instance, a
mutation that limits the availability of a ligand that activates an
enzyme following a Hill equation with early saturation may have a very
high frequency of neutral or mildly deleterious mutations compared
with a similar enzyme with late saturation. Nevertheless, uniform
sampling in our study is still useful in illustrating the main
evolutionary ideas presented here, which all have to do with relative
severity of mutations rather than their absolute fitness.

\begin{figure}[!htb]
\centering
  \includegraphics[width=0.85\textwidth]{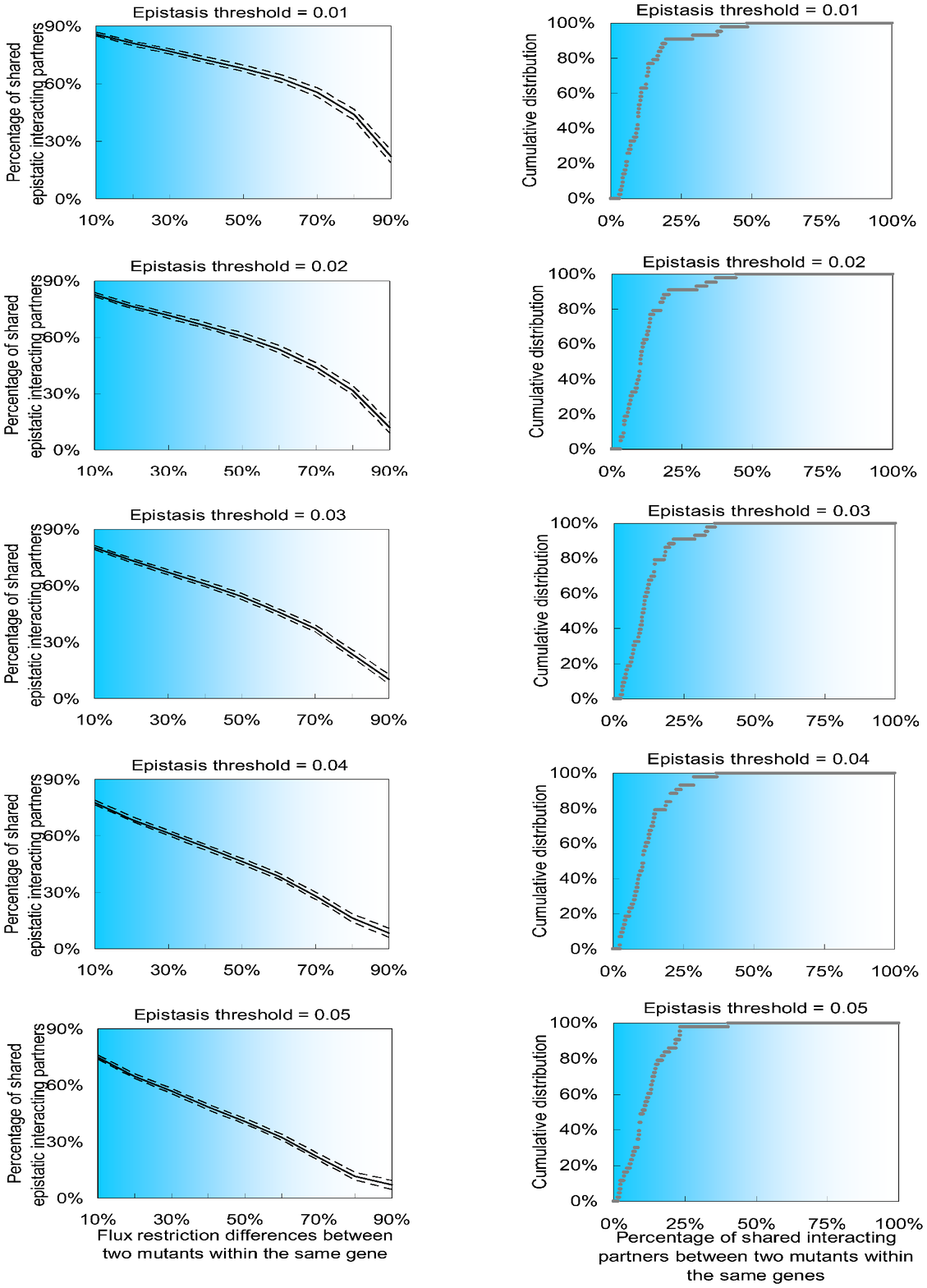}
  \caption{The conclusion that epistatic relations between genes are
allele-specific is robust to various epistasis thresholds. Left 5
panels: The FBA simulation results for the distribution of the
percentage of shared epistatic interaction partners between two mutant
alleles within the same gene. Solid and broken lines represent mean
and 95\% confidence intervals, respectively. Right 5 panels: The
cumulative distribution for the percentage of shared epistatic
interaction partners between two mutant alleles within the same gene
based on real experimental data. Both experimental and simulated
results are robust under various epistasis thresholds.}
\label{fig:sgeS4}

\end{figure}

\begin{figure}[!htb]
\centering
  \includegraphics[width=0.9\textwidth]{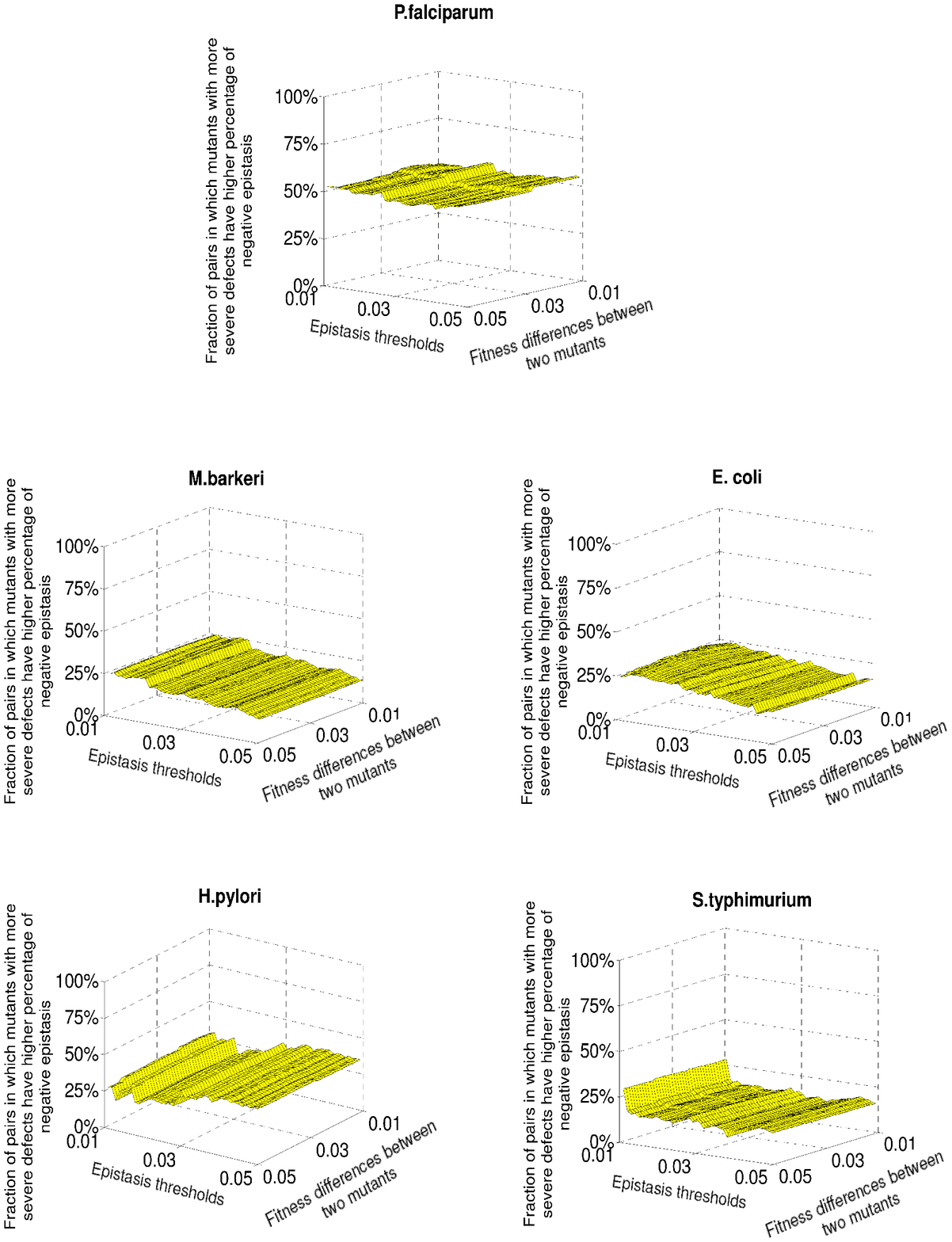}
\caption{The conclusion in Fig. 3, for which mutant alleles with more
severe defects tend to have a higher percentage of negative epistasis
in eukaryotes than bacteria and archaea, is robust under various
epistasis and fitness difference thresholds. The same methods to
generate Fig. 2C for S. cerevisiae are used here for the other five
species.}
\label{fig:sgeS5}
\end{figure}

\begin{figure}[!htb]
\centering
  \includegraphics[width=\textwidth]{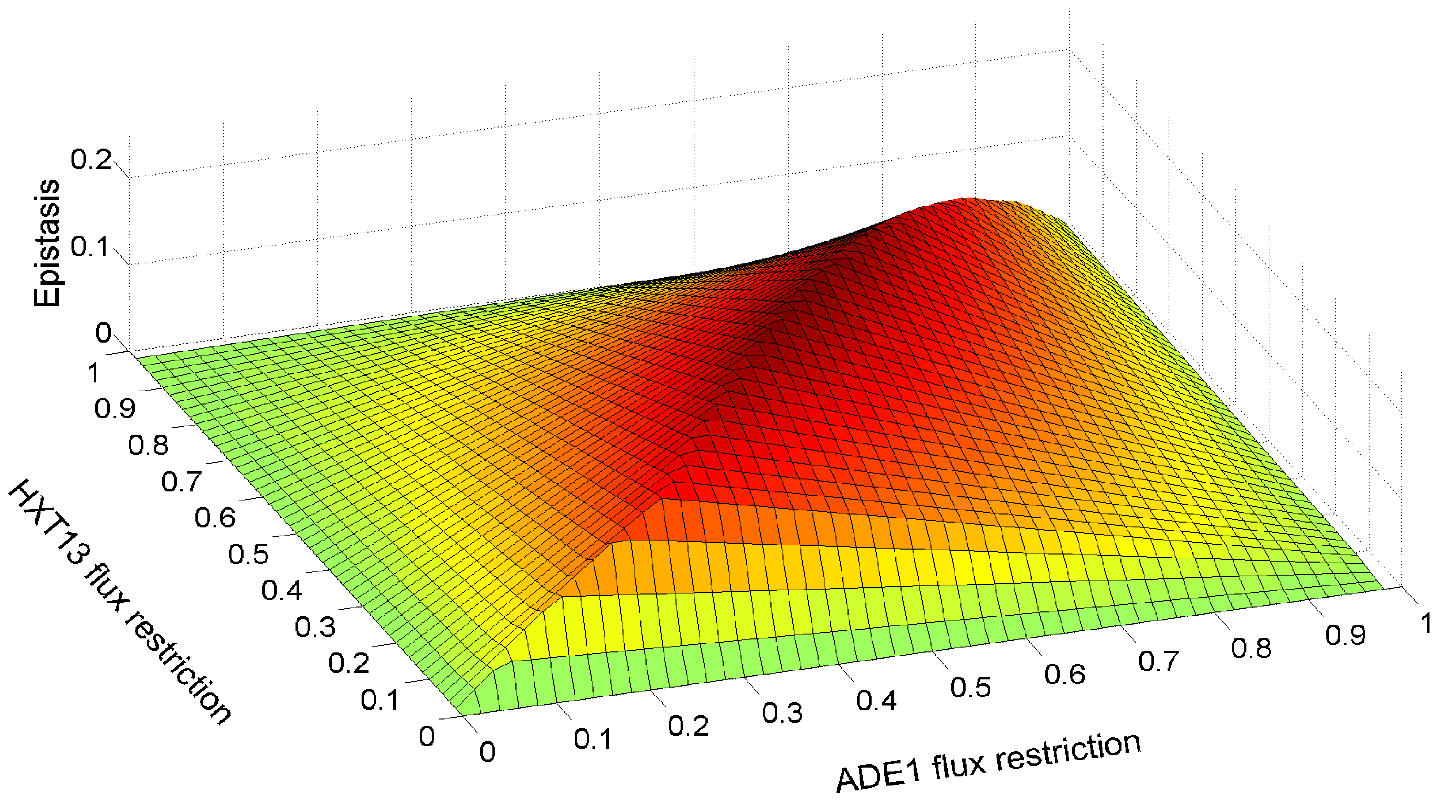}
\caption{An epistatic landscape exhibits smooth change in epistasis as
a function of the flux restriction for the genes HXT13 and ADE1. The
color corresponds to the z-axis (epistasis), with red being more
positive, and green being near zero. See dataset S3 for simulated
data. HXT13 is a hexose transporter and ADE1 is required for de novo
purine biosynthesis. The epistasis surface for HXT13 and ADE1 is quite
smooth, which is a fairly common pattern and we may infer that
epistasis, at least in metabolism, is often dependent on thresholds.}
\label{fig:sgeS6}
\end{figure}

\begin{figure}[!htb]
\centering
  \includegraphics[width=\textwidth]{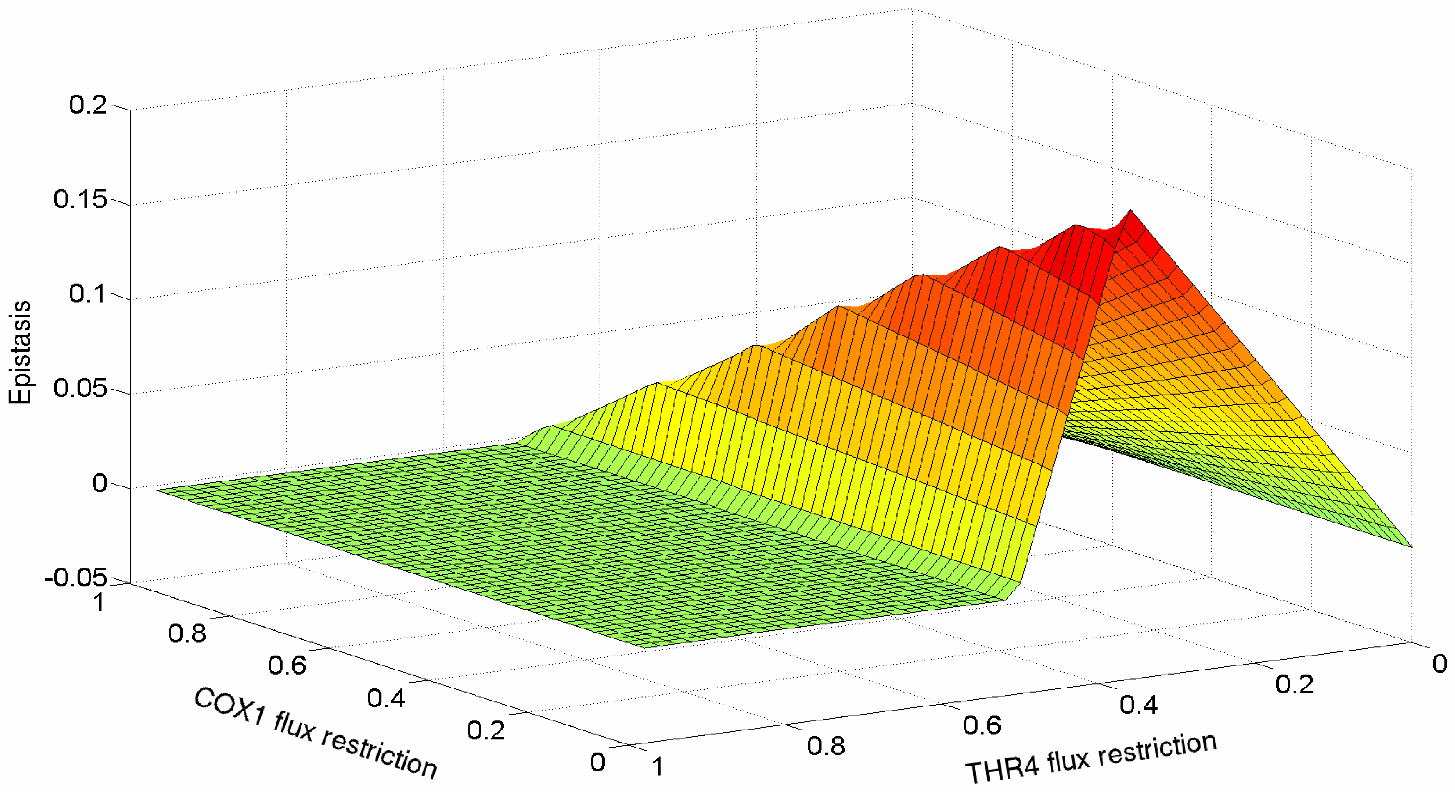}
\caption{An epistatic landscape exhibits a sharp transition to zero
epistasis, primarily as a consequence of the THR4 flux
restriction. The color corresponds to the z-axis (epistasis), with red
being more positive, and green being near zero. See dataset S4 for
simulated data. Epistasis is examined between threonine synthase gene
THR4 and COX1 (subunit 1 of cytochrome c oxidase). Both genes are
associated with mutually exclusive reactions. As shown in the figure,
there are regions where the epistasis is effectively zero (on the
order of 10-5) where the THR4 single mutant growth rate has only
changed very slightly, effectively allowing the mutations to act
independently. Once the THR4 mutant becomes more severe, the effects
are no longer independent.}
\label{fig:sgeS7}
\end{figure}

Third, measuring the presence of epistasis is subject to a choice of
threshold. Does the flux smoothly influence epistasis, or can
epistasis abruptly change or become zero? We have seen evidence of
both trends in our simulations. Though there are many different trends
in the magnitude of epistasis that we are currently investigating, we
present two cases to explore this issue (\Fig s~\ref{fig:sgeS6} and
\ref{fig:sgeS7} and Datasets S3 and S4). However, based on
\Fig~\ref{fig:sgeS4} and \ref{fig:sgeS5}, we have confirmed that our
major results are robust to a variety of epistasis thresholds. As
a result, although the choice of thresholds is a common problem for
research on epistasis, we are still confident that our conclusion is
unlikely to be significantly influenced by this factor. With these
limitations in mind, our observations identified several important
features for the epistasis among genes, and call on future
experimental and theoretical efforts to revisit genetics and
evolutionary theories that can integrate epistatic dynamics among
genes in biological systems.

\section{Methods}
\subsection{Experimental Dataset}
The experimental data were extracted from a global survey for the
epistatic interactions among genes in \textit{S. cerevisiae} \citep{Costanzo2010}. In
this original SGA study, the authors screened 1,712 \textit{S. cerevisiae}
query gene mutants against 3,885 array gene mutants to generate a
total of more than 5 million gene mutant pairs spanning all biological
processes. In each gene mutant pair, the epistasis value is calculated
based on the equation $\epsilon = W_{xy} - W_xW_y$, in which $W_{xy}$
is the fitness of an organism with two mutations in genes X and Y, and
$W_x$ or $W_y$ refers to the fitness of the organism with mutation
only at gene X or Y, respectively. In addition, a statistical
confidence measure ($p$-value) was assigned to each interaction based on
the observed variation of each double mutant across four experimental
replicates and estimates of the background error distributions for the
corresponding query and array mutants. Finally, a defined confidence
threshold ($\left|\epsilon\right| \geq 0.01$, P $<$ 0.05) was applied
to generate epistatic interactions \citep{Costanzo2010}.

\subsection{Flux Balance Analysis}

FBA frames the stoichiometric equations that describe the biological
reactions of a system as the following matrix equations, which is
possible because stoichiometric equations are linear \citep{Mo2009,
Becker2007, Smallbone2009a}.

\begin{equation}
\centering
\begin{tabular}{rl}
maximize   & $\mathbf{c}^T\mathbf{v}$                                     \\
subject to & $\mathbf{S v} = \frac{\D{\mathbf{x}}}{\D{t}} = \mathbf{0}$   \\
           & $\mathbf{v}_{lb} \preceq \mathbf{v} \preceq \mathbf{v}_{ub}$ \\
\end{tabular}
\end{equation}

The vector of concentration change over time
($\frac{\D{\mathbf{x}}}{\D{t}}$) is found by multiplying the
stoichiometric matrix $\mathbf{S}$ by a flux vector
$\mathbf{v}$. $\mathbf{S}$ has columns corresponding to each reaction
in the system, and rows corresponding to metabolites. Typically, one
or more enzymes correspond to each reaction, which allows us to see
how a genetic perturbation, such as a knockout, may affect the
system. The vector $\mathbf{v}$ consists of reaction fluxes and is
subject to upper and lower bounds $\mathbf{v}_{ub} = \left(u_1,u_2
\ldots, u_n\right)^T$ and $\mathbf{v}_{lb} = \left(l_1,l_2 \ldots,
l_n\right)^T$.  If we want to simulate the knockout or knockdown of an
enzyme, the fluxes corresponding to that enzyme can be constrained to
be zero or lower than WT, respectively. It is assumed that the change
in concentration over time is at steady state, therefore
$\frac{\D{\mathbf{x}}}{\D{t}} = \mathbf{0}$ in the FBA simulation
\citep{Smallbone2009a}.

The linear objective is written in terms of the $v_i$ with weight
coefficients $c_i$. Modified versions of COBRA and COBRA2 scripts,
popular FBA software packages written for MATLAB, were used to
implement our simulation framework \citep{Becker2007}. The method for calculating a
realistic WT flux for a given environment and organism model is taken
from \citealt{Smallbone2009a}. This method, termed geometric FBA,
attempts to choose a flux vector that is close to the average of all
optimal flux vectors. The geometric FBA solution is also a minimal
$\textbf{\textit{L}}^1$-norm solution, which has been previously
heralded as a good choice because it minimizes the total amount of
flux needed to achieve the objective, based on the fact that cells
would avoid having much unnecessary flux and wasted energy \citep{Smallbone2009a}. A
minimal $\textbf{\textit{L}}^1$-norm solution is advantageous in this
study because restricting fluxes for mutants based on unnecessarily
large WT fluxes may not constrain the system. Finally, the minimal
$\textbf{\textit{L}}^1$-norm solution avoids the problem of having
futile cycles, which are thermodynamically infeasible \citep{Smallbone2009a}.

Mutations of genes are simulated by the use of gene-reaction mapping
and flux constraints. Enzymes may be involved in multiple reactions
(i.e., pleiotropy). Although we often have Boolean rules describing
the relationship between genes in an enzyme complex, it is currently
extremely difficult to ascertain the exact contribution of each enzyme
to each reaction \citep{Banta2007}. Choosing the simplest unbiased approach, we
used gene-reaction mapping and uniformly constrained the flux through
each reaction associated to the gene being mutated. With one notable
exception \citep{He2010}, most research relating to simulation of
mutations with FBA has focused on null mutants \citep{Papp2004,
Ibarra2002, Harrison2007, Deutscher2006, Segre2005, Edwards2001,
Segre2002_sb2013}. Our simulation
approach, though simplifying the actual dynamics that result in
decreased fluxes in vivo, allows us to see behavior that was not
previously possible. To be consistent, we used the same equation and
threshold ($\left|\epsilon\right| \geq 0.01$) to calculate epistasis
for FBA results as we did for the experimental data.

\subsection{Population Genetics Model} 
\begin{figure}[!htb]
\centering
  \includegraphics[width=0.9\textwidth]{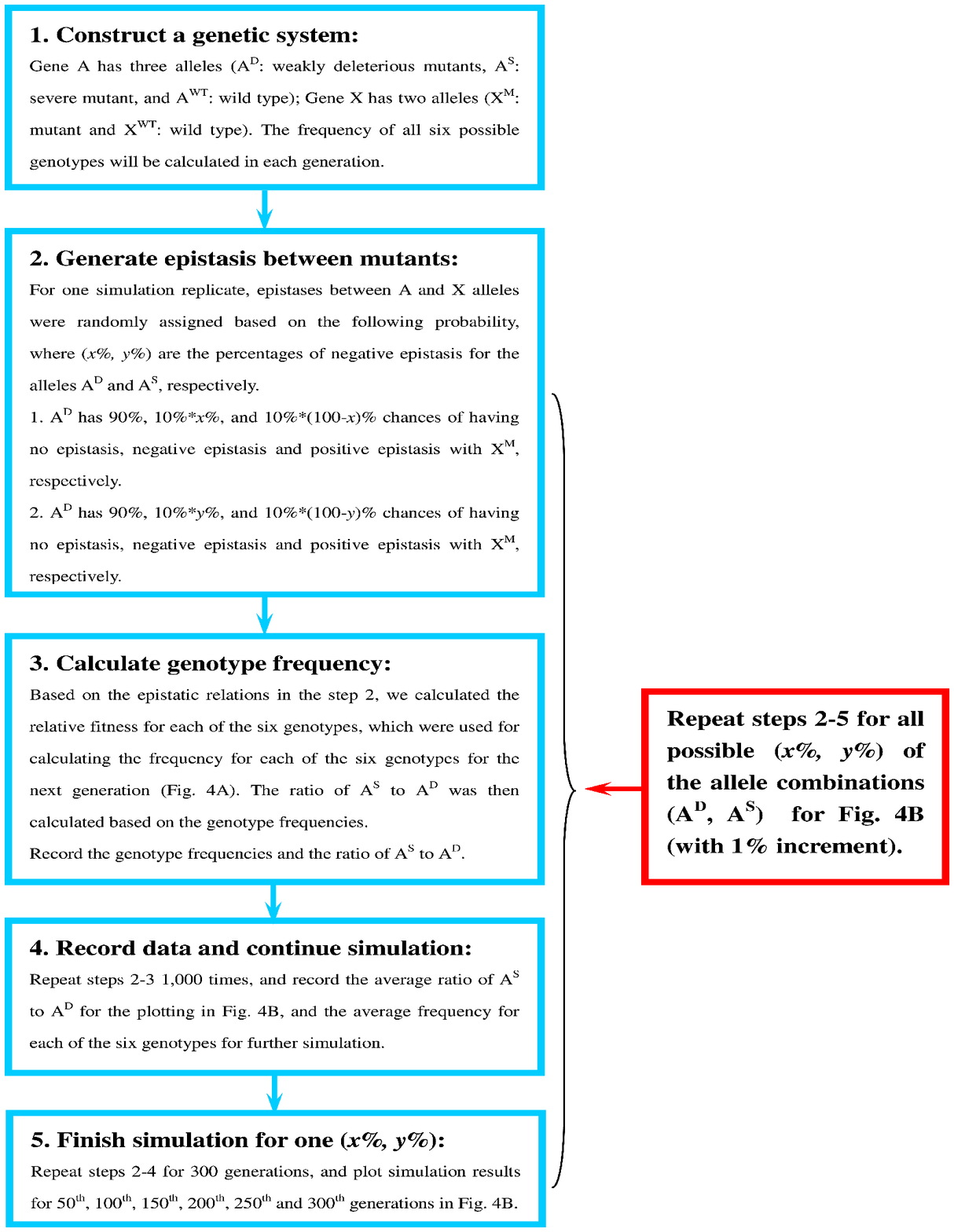}
\caption{A flow chart to illustrate the simulation process that
generates Fig. 4B. This procedure included 5 steps as indicated in the
5 blue boxes, and we have repeated step 2 to step 5 in the simulation
to produce all possible allele combinations, as highlighted in the red
box.}
\label{fig:sgeS8}
\end{figure}

A flowchart in
\Fig~\ref{fig:sgeS8} provides more illustration of the simulation
procedure. We constructed a
genetic system with a query gene A, which contains three different
alleles (A\textsuperscript{S}: severe mutant; A\textsuperscript{D}:
weakly deleterious mutants; and A\textsuperscript{WT}: WT) and a gene
X that has two different alleles (X\textsuperscript{M}: mutant and
X\textsuperscript{WT}: WT). The table in \Fig~\ref{fig:popgenPurging}A
explains how genotype
frequencies could be calculated from generation $T$ to generation
$T+1$ under natural selection. In the figure, $\rho$ and $\omega$
represent allele frequency and fitness, respectively. The average
fitness in generation $T$ could be calculated \citep{Hartl2007}.  We
simulated the ratio of allele frequency for the severe
(A\textsuperscript{S}) to the weak (A\textsuperscript{D}) mutant
alleles of the A gene under all possible combinations of the
percentages of negative epistasis for these two alleles, as shown on
the $x$ and $y$ axis of \Fig~\ref{fig:popgenPurging}B. For each
possible combination in each generation (a specific location on each
panel of \Fig~\ref{fig:popgenPurging}B), the following two-step
procedure was repeated 1,000
times. First, the epistatic relations (negative, positive, and no
epistasis) between the mutant alleles of the genes A and X were
randomly determined as the following: either A allele is assumed to
have 10\% possibility of having epistasis (either positive or
negative) with the allele X\textsuperscript{M} \citep{Costanzo2010}; when A and X
alleles do have epistasis, the likelihoods for the epistasis being
negative (and the remaining epistases are positive) are assigned
independently for A\textsuperscript{S} and A\textsuperscript{D}
alleles according to their location on
\Fig~\ref{fig:popgenPurging}B. Second, the fitness of each genotype
was calculated, which was then used to infer the genotype frequencies
in the next generation according to \Fig~\ref{fig:popgenPurging}A. The
average genotype frequencies among 1,000 randomizations
were then recorded for simulations in the next generation. The ratio
of allele frequency for the severe to the weak mutant alleles of the A
gene in each generation was calculated based on genotype frequencies
in that generation.

To make the simulation simple, the initial allele frequencies for the
severe, weak, and WT alleles of the A gene were assumed to be equal
(one-third), and the initial allele frequencies for the mutant and WT
of the X gene were also assumed to be equal (one-half). The fitness
was assumed to be 1, 0.99, and 0.98 for the WT, weak, and severe
mutant alleles of gene A, respectively, and 1 and 0.99 for the WT and
the mutant alleles of gene X, respectively. The positive and negative
epistasis values between A and X gene mutants were assumed to be 0.01
and −0.01, respectively. A variety of fitness differences between the
severe and weak alleles and epistasis values have also been used in
the simulations, and the trend remains the same.

\section{Acknowledgments}

We thank Dr. Ricardo Azevedo for his insights and critical comments on
the paper; Dr. Huifeng Jiang and Mr. Kaixiong Ye for discussion; and
the editor and two anonymous reviewers for constructive comments. This
work was supported by a startup fund from Cornell University, National
Science Foundation Grant DEB-0949556, and National Institutes of
Health Grant 1R01AI085286 (to Z.G.).  %

\section{Supporting Figures}

\section{Supporting Data}

Supporting datasets S1-S4 are available online (\texttt{DOI:
10.1073/pnas.1121507109}).  %

\end{document}